\documentclass[aps,twocolumn,groupedaddress]{revtex4}

\usepackage{epsfig}
\usepackage{amsmath}
\usepackage{braket}
\usepackage{xspace}
\usepackage{graphicx}
\usepackage{lineno}

\begin{document}
\draft

\title{Classification of coherent peaks in two-terminal quantum devices into normal and anomalous Kondo peaks}
\author{Jongbae Hong}
\affiliation{School of Physics and Astronomy, Seoul National University, 1 Gwanak-ro, Gwanak-gu, Seoul 08826, Korea}
\date{\today}

\begin{abstract}
Coherent peaks arising in the differential conductance of quantum dot (QD) and quantum point contact (QPC) devices are classified into two categories—normal and anomalous Kondo peaks—according to the underlying spin dynamics and the form of the scaling function to which the scaled temperature-dependent linear conductance collapses. The zero-bias peaks (ZBPs) observed in QPCs and in the triplet state of the even sector of quantum dot single-electron transistors (QDSETs) are identified as normal Kondo peaks, formed by spin dynamics involving spin exchange, a symbolic characteristic of the Kondo effect. For these ZBPs, the scaling temperature coincides with half the full width at half maximum (FWHM). In contrast, the ZBP observed in the odd sector of QDSETs and all finite-bias coherent peaks—including the coherent side peaks of QPCs and the split ZBP in the singlet state of the QDSET even sector—are identified as anomalous Kondo peaks, because they arise from spin dynamics without spin exchange, and their scaling temperature does not coincide with half the FWHM. To support these findings, we reproduce gate-voltage-dependent differential conductance line shapes measured in the odd sector of a QDSET, demonstrating that its ZBP originates from a combination of two coherent side peaks explicitly observed in QPCs.
\end{abstract}


\maketitle \narrowtext 

\section{Introduction}
\begin{figure}[b] 
\centering
\includegraphics[width=2.7 in]{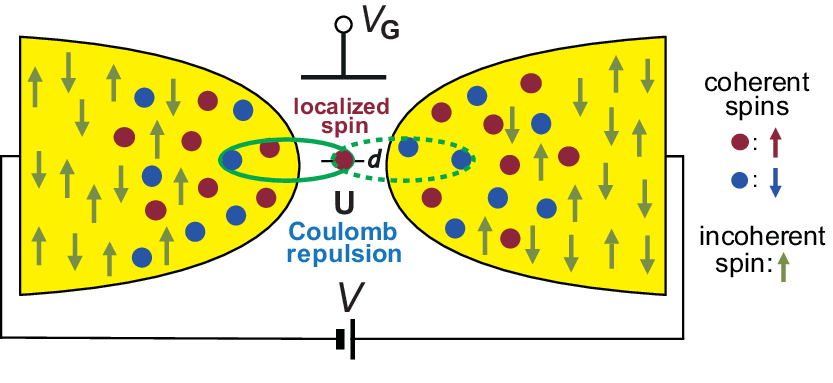}
\caption{Schematic illustration of a two-terminal nanodevice in which transport occurs via entangled-state tunneling formed by a linear combination of two Kondo singlets. $U$ denotes the Coulomb repulsion at the position of the localized spin and $V_{\rm G}$ the side-gate voltage. 
Solid dots represent the coherent spins at the Fermi levels of the reservoirs.}
\label{fig1}
\end{figure}

Quantum technology has become one of the most prominent global research directions in modern condensed matter and device physics. 
Advances in nanofabrication techniques have enabled the realization of confined nanostructures such as quantum dots (QDs) and quantum point contacts (QPCs). 
The basic design of these devices consists of two metallic reservoirs and a localized spin between them, as illustrated in figure~\ref{fig1}. 
Extensive research on quantum nanoscopic devices exhibiting the so-called Kondo peak has been conducted since the late 1990s. 

The localized spin may exist in an artificially created atom (the QD)~\cite{Ashoori,Goldhaber,Cronenwett98} or emerge spontaneously in the narrow constriction of a QPC~\cite{Rejec,Yoon,Bauer}. 
At sufficiently low temperatures, a Kondo singlet forms between the localized spin and an itinerant electron in the reservoirs. 
The first scaling analysis for nanoelectronic devices was performed on quantum-dot single-electron transistors (QDSETs)~\cite{Goldhaber-PRL}, followed by a corresponding analysis for QPCs~\cite{Cronenwett}. 
These studies demonstrated that the low-energy transport properties of both QDs and QPCs are governed by Kondo physics, as evidenced by the collapse of temperature-dependent linear conductance data at various gate voltages onto a universal scaling function when temperature is scaled by the appropriate characteristic temperature. 
It is worth noting, however, that the scaling functions employed for the QDSET and QPC zero-bias peaks (ZBPs) differ.

The fundamental observable that carries the dynamics of the system is the nonlinear differential conductance $dI/dV$ as a function of the bias voltage $V$, where $I$ is the source–drain current. 
In QDSETs, the measured $dI/dV$ line shapes typically show a ZBP and two broad incoherent Coulomb peaks when the gate voltage lies in the odd sector of the Coulomb diamond formed from the source-drain bias versus gate voltage plot. 
In contrast, QPCs exhibit a relatively narrow and small ZBP, two pronounced coherent side peaks, and two broad Coulomb peaks that are often not seen in experimental data~\cite{Cronenwett,Sarkozy,Chen,Ren}.

Later studies on the even sector of QDSETs revealed a split ZBP in the singlet state and a ZBP accompanied by two coherent side peaks in the triplet state~\cite{Roch}. 
The split peak has been interpreted in terms of the two-stage Kondo effect~\cite{Roch,Hofstetter,Florens,Petit,Karki,Guo}. 

Scaling analysis is a crucial tool in studies of Kondo systems~\cite{Scaling}. 
For QD and QPC devices, scaling is usually performed for temperature-dependent linear conductance at zero bias. 
In the present work, however, we extend the scaling analysis to finite-bias coherent peaks, which have generally been excluded from previous analyses. 
Including these peaks allows us to classify all coherent peaks according to their respective scaling functions.

There are two basic types of scaling function:
\begin{equation}
G_{\rm I}(T)=G_0[1+(2^{1/s}-1)(T/T_{\rm AK})^2]^{-s},
\end{equation}
with $s=0.22$, where $G_0$ is the zero-temperature conductance; and
\begin{equation}
G_{\rm II}(T)/(2e^2/h)=0.5[1+(2^{1/s}-1)(T/T_{\rm NK})^2]^{-s}+0.5.
\end{equation}
The distinction between the two scaling temperatures, $T_{\rm AK}$ and $T_{\rm NK}$, reflects fundamentally different underlying spin dynamics, etc., as discussed later. 

We exclude the scaling analysis for the QDSET odd-sector and QPC ZBPs, since their scaling behavior—$G_{\rm I}(T)$ for the former~\cite{Goldhaber-PRL,Petit,Grobis,Wiel,Heersche} and $G_{\rm II}(T)$ for the latter~\cite{Cronenwett,Chen}—is well established. 
Our aim is instead to perform scaling analysis for various finite-bias coherent peaks, together with the ZBP of reference~\cite{Roch}, observed in the triplet state of the QDSET even sector. 
We also identify the fundamental physical factors that differentiate the two scaling types. 
To this end, we theoretically reproduce the measured differential conductance line shapes in the QDSET odd sector. 

The remainder of this paper is organised as follows. 
Section~\ref{sec:scaling} presents the results of scaling analysis for various coherent peaks in QD and QPC devices and classifies them according to the relevant scaling functions. 
Section~\ref{sec:formula} briefly introduces the Green’s function technique in operator space and derives a convenient differential conductance formula. 
Section~\ref{sec:reproduct} reproduces the experimental differential conductance line shapes to elucidate the origin of the ZBP in the QDSET odd sector. 
Section~\ref{sec:discussion} provides evidence supporting the identification of $T_{\rm AK}$ and $T_{\rm NK}$ as anomalous and normal Kondo temperatures, respectively. 
Section~\ref{sec:conclusion} summarizes our conclusions, followed by detailed appendices.

\begin{figure}[t] 
\centering
\includegraphics[width=3.0 in]{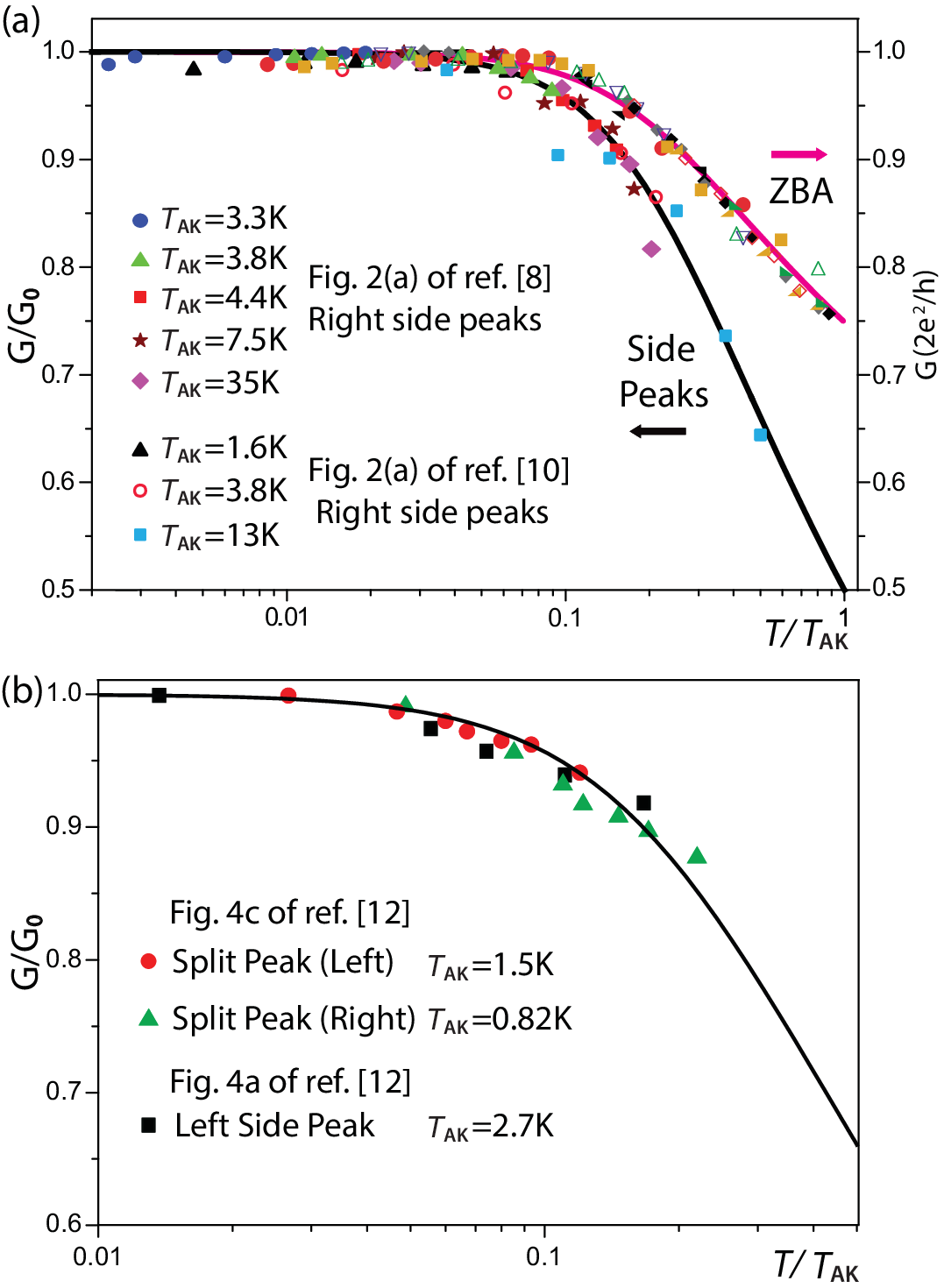}
\caption{(a) Temperature-dependent finite-bias linear conductance scaled by $T_{\rm AK}$ for different gate voltages, plotted as a function of 
$T/T_{\rm AK}$. The scaling behavior of the zero-bias anomaly from reference~\cite{Cronenwett}, scaled by $T_{\rm NK}$, is shown for comparison. 
The black and magenta curves represent $G_{\rm I}(T)$ and $G_{\rm II}(T)$, respectively.
Data sources and $T_{\rm AK}$ values are provided.
(b) Same as (a), but for data extracted from line shapes at the singlet (red circles and green triangles) and triplet (dark squares) states in the QDSET even sector~\cite{Roch}.}
\label{fig2}
\end{figure}

\section{Scaling analysis for differential conductance maxima} \label{sec:scaling} 

Scaling analysis is a powerful method for identifying Kondo effects in mesoscopic systems. 
Here, we extend this analysis to the temperature dependence of the differential conductance maxima associated with various coherent peaks, including both zero- and finite-bias features. 
As mentioned above, we focus on the finite-bias coherent peaks observed in the singlet and triplet states of the QDSET even sector, as well as the side peaks of the QPC device.

The scaling results for the finite-bias coherent peaks are presented in figure~\ref{fig2}: Figure~\ref{fig2}(a) corresponds to the QPC device~\cite{Cronenwett,Chen}, while figure~\ref{fig2}(b) corresponds to the singlet and triplet states of the QDSET even sector~\cite{Roch}. 
For comparison, in figure~\ref{fig2}(a) we include the scaled zero-bias anomaly data from reference~\cite{Cronenwett}, which are scaled using the function 
$G_{\rm II}(T)$. 
We conclude that all temperature-dependent finite-bias linear conductance data scale onto the scaling function $G_{\rm I}(T)$, regardless of the type of nanodevice.

\begin{figure}[t] 
\centering
\includegraphics[width=2.7 in]{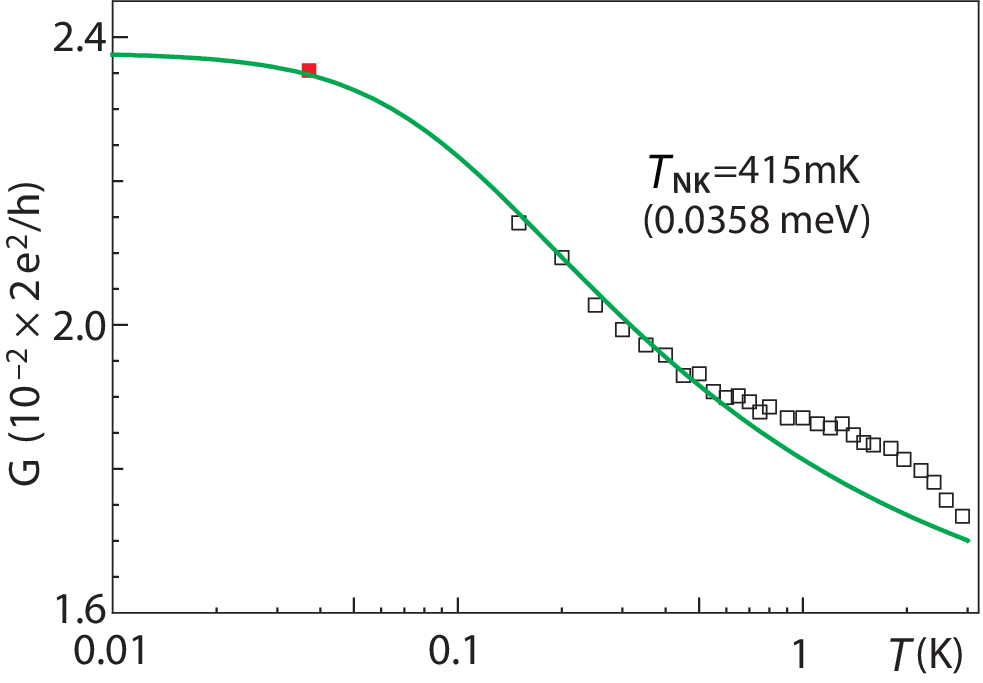}
\caption{The curve of equation~(\ref{Cond-scale}) (green line) and temperature-dependent linear conductance data (open squares) from Fig. 4b of reference~\cite{Roch}. 
The red square represents the linear conductance at 37 mK from Fig. 4a of reference~\cite{Roch}.}
\label{fig3}
\end{figure}

An interesting remaining case in the scaling analysis is the zero-bias anomaly generated in the triplet state of the QDSET even sector~\cite{Roch}. 
Since the scaling function for the zero-bias conductance of the QDSET odd sector is known to be $G_{\rm I}(T)$, it is intriguing to determine which scaling function applies to the zero-bias conductance of the QDSET even sector.

Remarkably, as shown in figure~\ref{fig3}, the temperature-dependent linear conductance taken from Fig. 4b of reference~\cite{Roch} scales well onto
\begin{equation} 
G(T)=0.01718\, G_{\rm II}(T)+0.0066\,(2e^2/h),
\label{Cond-scale}
\end{equation}
with a scaling temperature of $T_{\rm NK}=415$ mK, which corresponds to half the full width at half maximum (FWHM), 0.0358 meV, taken from of the lowest-temperature ZBP in Fig. 4a of reference~\cite{Roch}. 
In figure~\ref{fig3}, the tendency of the data to deviate from the scaling function above $T_{\rm NK}$ is similar to that observed in previous studies~\cite{Goldhaber-PRL, Cronenwett, Wiel}. 

The second term in equation~(\ref{Cond-scale}) represents the background conductance, while the small coefficient of the first term is attributed to the low conductance amplitude. 
Explaining the origin of this low amplitude is beyond the scope of this work; this issue may be clarified by reproducing the differential conductance line shapes in future studies.

A qualitative description of the differential conductance for both singlet and triplet states in the QDSET even sector has been provided within the non-crossing approximation~\cite{Roura1,Roura2}, which, however, has limitations in describing ground-state physics. 
The authors considered only $G_{\rm I}(T)$ as the scaling function, even for the zero-bias anomaly in the triplet state—an unfortunate oversight. 
For a quantitative description, more advanced dynamical methods may be required to treat entangled-state tunneling through singlet and triplet states in QDSETs.

The results of the scaling analysis can be summarized as follows:  \\
\noindent (A) In the odd sector of the QDSET, the ZBP is  \\
\-\hspace{0.5cm} scaled by the function $G_{\rm I}(T)$; \\
(B)  All finite-bias coherent peaks, including those of   \\ 
\-\hspace{0.5cm} the singlet and triplet states in the QDSET even  \\
\-\hspace{0.5cm} sector, scale onto $G_{\rm I}(T)$; \\
(C) The zero-bias anomaly of the QPC device scales \\
\-\hspace{0.5cm} onto $G_{\rm II}(T)$; \\
(D) The zero-bias anomaly in the triplet state of the \\
\-\hspace{0.5cm} QDSET scales onto $G_{\rm II}(T)$.\\
In other words, for zero-bias anomalies, those accompanied by two coherent side peaks scale onto $G_{\rm II}(T)$, while ZBPs without coherent side peaks and 
finite-bias coherent peaks scale onto $G_{\rm I}(T)$.

One of the most intriguing aspects of the present scaling results is that the ZBP in the QDSET odd sector belongs to the same group as the finite-bias coherent peaks. 
This apparent inconsistency can be resolved by considering the spectral line shapes and the structural characteristics of the nanodevices.

As illustrated schematically in figure~\ref{fig1}, both QD and QPC devices share a similar structure. 
The left–right entanglement responsible for generating two coherent side peaks is thus expected in both systems, implying that such side peaks must also exist in the QDSET. 
Therefore, the above issue can be addressed by identifying the location of the coherent side peaks within the differential conductance of the QDSET odd sector. 
To achieve this, a theoretical reproduction of the measured differential conductance of the QDSET is required. 
In Section~\ref{sec:reproduct}, we perform this reproduction using the methods described in Section~\ref{sec:formula}.

\section{Differential conductance formula} \label{sec:formula} 

To reproduce the measured differential conductance obtained for a QDSET, we employ the same theoretical framework used in our previous work on QPC devices~\cite{iop-qpc}. 
The gate-voltage-dependent differential conductance, $\frac{dI}{dV}(V_{\rm G})$, where $I$ is the source–drain current and $V$ the applied bias, is expressed as:
\begin{equation} 
\frac{dI}{dV}(V_{\rm G}) = \left. \frac{2e^2}{h} \widetilde{\Gamma}^{\rm sn}(V_{\rm G}) {\rm Im} {\mathcal{G}}_{dd\uparrow}^{+\rm sn}(\omega, V_{\rm G}) \right|_{\hbar\omega = eV},
\label{Conductance}
\end{equation}  
where $\widetilde{\Gamma}^{\rm sn}(V_{\rm G})$ is the effective coupling function in the two-terminal device (see figure~\ref{fig1})~\cite{Meir}, and 
${\rm Im} \, {\mathcal{G}}_{dd\uparrow}^{+\rm sn}(\omega, V_{\rm G})$ denotes the imaginary part of the on-site retarded Green’s function at the localized spin site labeled by ${\it d}$. 
The superscript ``sn'' refers to the steady-state nonequilibrium condition. 

The retarded Green’s function $\mathcal{G}_{dd\uparrow}^{+\rm sn}(\omega)$ corresponds to the $dd$ element of the Green’s function matrix:
\begin{equation} 
i\mathcal{G}_{dd\uparrow}^{+\rm sn}(\omega) = \left[\frac{1}{z \mathbf{I} + i \mathbf{L}}\right]_{dd},
\end{equation} 
where $z = -i\omega + 0^+$, $\mathbf{I}$ is the identity operator, and $\mathbf{L}$ is the Liouville operator defined in Liouville space~\cite{Fulde}. 
The relation $-{\rm Im}{\mathcal G}^{+}_{dd\uparrow}(\omega)={\rm Re}({\bf M}^{-1})_{dd}$ holds, where $\mathbf{M} = z \mathbf{I} + i \mathbf{L}$.

\begin{figure}[t] 
\centering
\includegraphics[width=2.7 in]{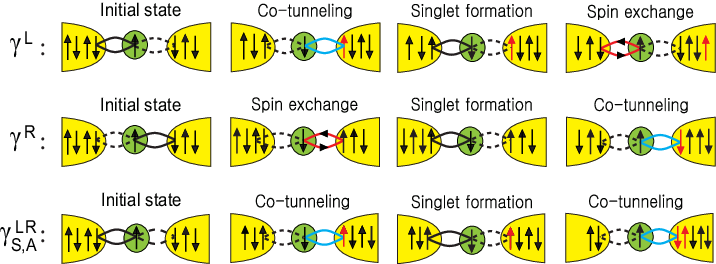}
 \caption{Spin dynamics of unidirectional entangled-state tunneling with ($\gamma^{L,R}$) and without ($\gamma^{LR}_{S,A}$) spin exchange.}
\label{fig4}
\end{figure}

We construct the Liouville matrix $\mathbf{L}$ using basis operators that span the working Liouville space~\cite{Hong11} and obtain an equivalent $5 \times 5$ matrix $\mathbf{M}_r$ via matrix reduction~\cite{Lowdin,Mujica}:
\begin{eqnarray}
{\rm\bf M}_r=\left(\begin{array}{c c c c c} -i\omega & \gamma^L &
-U^L_{j^-} & \gamma^{LR}_S & \gamma^{LR}_A \\ -\gamma^L & -i\omega
& -U^L_{j^+} & \gamma^{LR}_A & \gamma^{LR}_S \\
U_{j^-}^{L*} &  U_{j^+}^{L*} & -i\omega &  U^{R*}_{j^+} &
U^{R*}_{j^-} \\  -\gamma^{LR}_S & -\gamma^{LR}_A & -U_{j^+}^R  &
-i\omega & -\gamma^R \\
 -\gamma^{LR}_A &  -\gamma^{LR}_S &  -U_{j^-}^R  & \gamma^R & -i\omega
\end{array}\right)+i{\bf \Sigma}, \nonumber \\
\label{reduced}
\end{eqnarray} 
where $\omega$ is defined as $\omega - \epsilon_d - U \langle n_{d\downarrow} \rangle$, with $\epsilon_d$, $U$, and $\langle n_{d\downarrow} \rangle$ representing the localized energy level, on-site Coulomb interaction, and spin-down occupancy, respectively.
A detailed derivation of equation~(\ref{reduced}) is provided in reference~\cite{iop-qpc}.
The matrix elements are given in appendix A, and the corresponding spin dynamics encoded in the $\gamma$ parameters are described in appendix B, and the pictorial descriptions are shown in figure~\ref{fig4}.

The matrix elements of the self-energy $i{\bf \Sigma}$ are given by 
$i{\bf \Sigma}_{pq}=\eta_{pq}[i{\bf \Sigma}_0^L(\omega)+i{\bf \Sigma}_0^R(\omega)]$, with coefficients $\eta_{pq}$ appeared during matrix reduction. 
Here,  $i{\bf \Sigma}_0^\nu(\omega)=\pi(\tilde{V}^\nu)^2\rho_0[\sqrt{1-(\omega^2/D^2)}+i\omega/D],$
is the self-energy of the noninteracting Anderson model with a wide semi-elliptic band $\rho_0(\omega)=\rho_0\sqrt{1-(\omega^2/D^2)}$.

The coupling function $\Gamma^\nu(\omega)$ is twice the imaginary part of ${\bf \Sigma}_0^\nu(\omega)$, i.e., 
$\Gamma^\nu(\omega)=2\pi(\tilde{V}^\nu)^2\rho_0(\omega)$.
Because the bandwidth $2D$ is sufficiently wide, $\Gamma^\nu(\omega)$ can be treated as a constant with respect to $\omega$, denoted by  
$\Gamma^\nu=2\pi(\tilde{V}^\nu)^2\rho_0=2\Delta^\nu$.
The parameter $\Delta^\nu$ depends on gate voltage $V_{\rm G}$ through the hybridization $\tilde{V}^\nu(V_{\rm G})$, while the sum $\Delta^L + \Delta^R$ remains constant. 
We adopt $\Delta=(\Delta^L + \Delta^R)/2$ as the energy unit throughout this study.

Finally, the convenient differential conductance formula for the two-terminal system in figure~\ref{fig1} is written as~\cite{iop-qpc}:
\begin{equation}
\frac{dI}{dV}= \left. \frac{2e^2}{h} \widetilde{\Gamma}^{\rm sn}(V_{\rm G}){\rm Re}({\rm\bf M}_r^{-1})_{33}\right|_{\hbar\tilde{\omega}=eV}.
\label{eq:new-dIdV}
\end{equation}

\section{Reproduction of differential conductance} \label{sec:reproduct} 

\begin{figure}[t] 
\centering
\includegraphics[width=3.0 in]{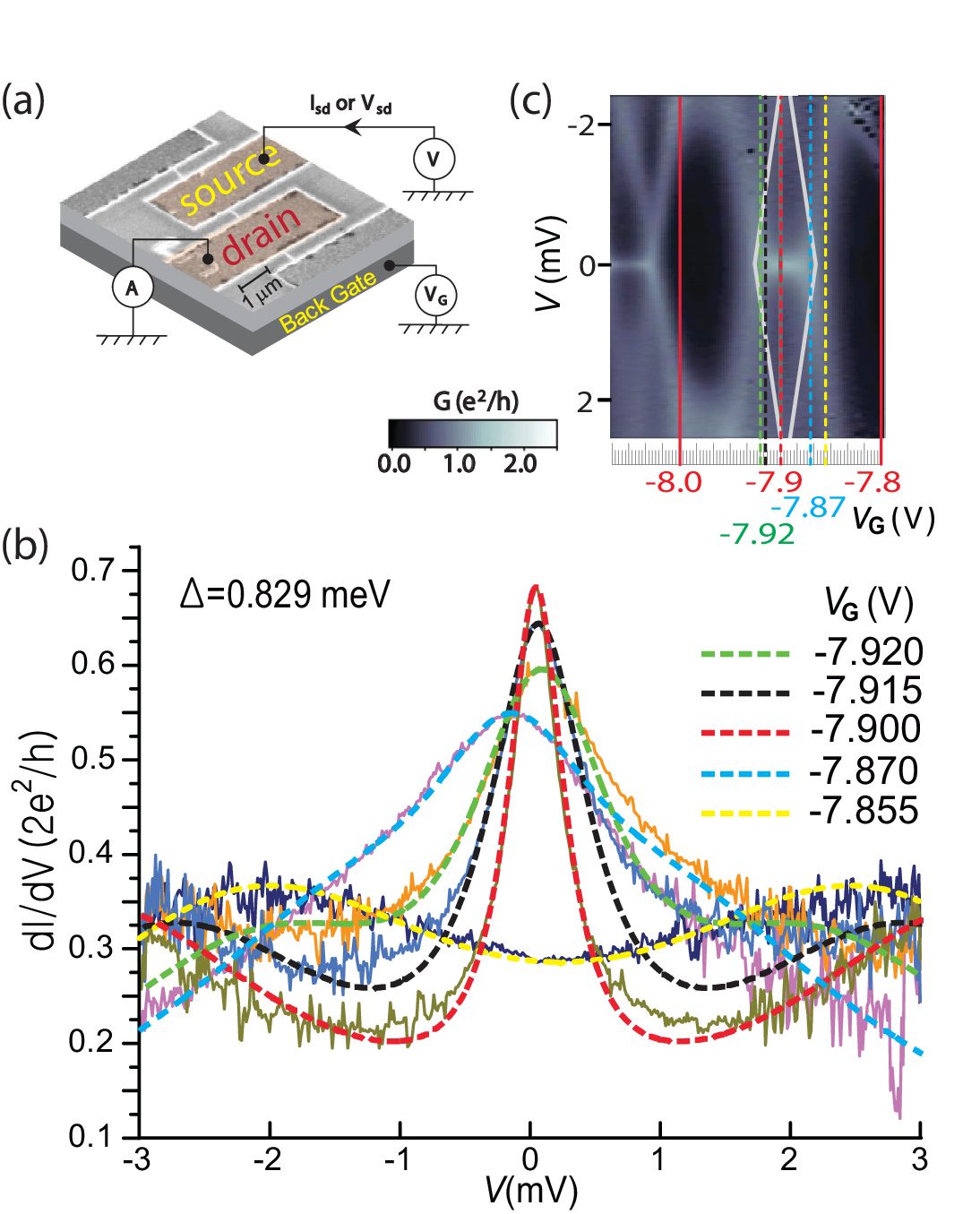}
 \caption{(a) Schematic of the CNT QDSET. 
A highly doped Si wafer and a SiO$_2$ layer are used as a back gate and an insulating barrier, respectively. 
The CNT channel length is designed to be 300 nm with a diameter of approximately 1.5 nm. 
(b) Experimental differential conductance curves (solid lines) and theoretically reproduced curves (dashed lines) fitted on the left side for several gate voltages.
Theoretical results are shifted to match the experimental peak positions. 
(c) Gray-scale plot of differential conductance as a function of source–drain bias $V$ and gate voltage $V_{\rm G}$.
}
\label{fig5}
\end{figure}

Experimentalists who fabricated a carbon nanotube (CNT) QDSET~\cite{kriss}, as shown in figure~\ref{fig5}(a), measured its differential conductance, and provided us their unpublished data. 
All measurements were performed at a base temperature below 100 mK in a dilution refrigerator. 
An external magnetic field of 1 kG was applied to the Al superconducting electrode to drive the system into a normal state.
They obtained the $V_{\rm G}$-dependent differential conductance (solid lines) and its gray-scale plot as a function of source–drain bias and gate voltage, as shown in figures~\ref{fig5}(b) and~\ref{fig5}(c), respectively. 

The observed line shapes consist of one ZBP and two broad Coulomb peaks, typical features of the QDSET odd sector~\cite{Wiel,Nygard}. 
These data are theoretically reproduced by determining the parameters in equation~(\ref{reduced}) together with $\widetilde{\Gamma}^{\rm sn}(V_{\rm G})$ in equation~(\ref{eq:new-dIdV}).

To determine the parameters, we employed the same self-energy coefficients $\eta_{pq}$ used in the previous study~\cite{iop-qpc}: 
$\eta_{11} = \eta_{15} = \eta_{55} = 0.253$, 
$\eta_{12} = \eta_{14} = \eta_{25} = \eta_{45} = 0.254$, 
$\eta_{22} = \eta_{24} = \eta_{44} = 0.260$, and 
$\eta_{33} = 1$, with symmetric indices. 
The prefactor $\widetilde{\Gamma}^{\rm sn}(V_{\rm G})$ was determined by matching the theoretical and experimental maximum values. 
Because the unidirectionality of the entangled-state tunneling imposes $\gamma^{LR}_{S} = \gamma^{LR}_{A}$, the remaining parameters in matrix ${\rm\bf M}_r$ are 
$\gamma^{L}$, $\gamma^{R}$, $\gamma^{LR}_{S,A}$, $U_{J^-}^{L}$, $U_{J^+}^{L}$, $U_{J^-}^{R}$, and $U_{J^+}^{R}$.
These can be further reduced using results from the atomic limit analysis described below.

The atomic limit analysis for large $U$ at half-filling (${\rm Im}[U_{j^\pm}^{L,R}] = 0$) provides the following insights~\cite{iop-qpc}: \\

\noindent (a) The spectral weight of the zero-bias anomaly is \\
\-\hspace{0.5cm}  governed by $\gamma^L$ and $\gamma^R$ as \\ 
\-\hspace{0.5cm} $8(\gamma^L\gamma^R)^2/[U^2\{(\gamma^L)^2+(\gamma^R)^2\}]$;\\
(b) The spectral weight of the coherent side peaks is \\
\-\hspace{0.5cm} governed by $\gamma^{LR}_{S,A}$ as $8(\gamma^{LR}_{S,A}/U)^2$;\\
(c) The positions of the two coherent side peaks are \\
\-\hspace{0.5cm} determined by $\gamma^L$ and $\gamma^R$ as $\pm\sqrt{[(\gamma^L)^2+(\gamma^R)^2]/2}$.\\

To obtain a single ZBP, we set $\gamma^L = \gamma^R = 0$ to eliminate the zero-bias anomaly—characteristic of QPCs— from the condition (a) above.
The condition (c) then causes two coherent side peaks to be pushed inward.
Furthermore, we find that setting $U_{j^-}^{L} = U_{j^+}^{L} = U_{j^-}^{R} = U_{j^+}^{R}$ yields a single ZBP without splitting. 
Now, there are only two parameters left: $\gamma^{LR}_{S,A}$ and $U_{j^\mp}^{L,R}$.

Next, we discuss the compatibility between the two conditions $\gamma^{L,R}=0$ and 
$U_{j^-}^{L}=U_{j^+}^{L}=U_{j^-}^{R}=U_{j^+}^{R}$. 
The parameters $\gamma^L$ and $\gamma^R$ describe unidirectional entangled-state tunneling, as illustrated in figure~\ref{fig4}.
Note that backward motion appears only in spin exchange.
The spin dynamics of $\gamma^{L,R}$ are detailed in appendix~B based on the operator dynamics given in equation~(\ref{gamma-L}); see reference~\cite{iop-qpc} for details.

In contrast, the parameters $U_{j^\mp}^{L,R}$ represent effective Coulomb interactions screened by the motion of incoherent spins flowing to or from the left and right reservoirs.
Hence, the condition $U_{j^-}^{L}=U_{j^+}^{L}=U_{j^-}^{R}=U_{j^+}^{R}$ restricts screening by backward motion on each side of the localized spin, since subscripts $j^\mp$ denote spin motions $\longrightarrow\mp\longleftarrow$. 
Excluding backward motion is equivalent to suppressing spin exchange, resulting in $\gamma^L=\gamma^R=0$. 
Therefore, the two conditions are consistent with each other.

Theoretical results (dashed lines) obtained with the parameters listed in Table~I are superimposed on the experimental data (solid lines) by aligning the peak positions, as shown in figure~\ref{fig5}(b). 
The values of $U_{j^\mp}^{L,R}$ in Table~I are consistent with the Coulomb diamond boundaries shown in figure~\ref{fig5}(c).
We find that $V_{\rm G} = -7.855$~V belongs to the even sector, where only Coulomb peaks are observed, whereas the other gate voltages belong to the odd sector.

Finally, we emphasize that the ZBP observed in the QDSET odd sector originates from the combination of two coherent side peaks.
This explains why the ZBP of the QDSET odd sector belongs to the same scaling group as the coherent side peaks, as shown in figure~\ref{fig2}.

\begin{table} [t]
\centering
\caption{\textbf{Variation of parameters with $V_{\rm G}$} }
   \vspace{0.3cm}
    \setlength{\tabcolsep}{2.7 pt}
   \begin{tabular}{c c c c c c}
\hline\hline \\ [-2ex] $V_{\rm G}$ & $\gamma^L$ & $\gamma^R$ & $\gamma^{LR}_{S,A}$ & $U_{J^\pm}^{L,R}$ & 
$\widetilde{\Gamma}^{\rm sn}$
\\ [0.5ex] \hline  
-7.920 & 0   & 0     &  0.72   & 0.72     &  0.5955    \\
-7.915 & 0   & 0     &  0.75   & 1.10     &  0.6445    \\
-7.900 & 0   & 0     &  0.67   & 1.48     &  0.6800    \\
-7.870 & 0   & 0     &  0.60   & 0.34     &  0.5480   \\
-7.855 & 0   & 0     &  0.00   & 1.10     &  0.6250    \\
 [0.5ex] \hline
\end{tabular}
\label{table}
\end{table}

\section{Normal Kondo peak vs Anomalous Kondo peak}\label{sec:discussion}

In this section, we discuss the two scaling temperatures, $T_{\rm AK}$ and $T_{\rm NK}$, and explain why we refer to them as the anomalous and normal Kondo temperatures, respectively.

In the previous section, we showed that the ZBP in the odd sector of the QDSET originates from the coherent side peaks explicitly appearing in the QPC device. 
Their spectral weight is determined by $\gamma^{LR}_{S,A}$, as demonstrated by the atomic limit analysis in Section~\ref{sec:reproduct}. The spin dynamics of $\gamma^{LR}_{S,A}$ correspond to consecutive singlet co-tunneling processes without spin exchange, as illustrated in figure~\ref{fig4}. In contrast, spin dynamics involving spin exchange (represented by $\gamma^{L,R}$ in figure~\ref{fig4}) generate the zero-bias anomaly in QPCs, as discussed in reference~\cite{iop-qpc}.  

Thus, we have clarified that the ZBPs appearing in the QDSET odd sector and in QPCs are fundamentally different in both spin dynamics and scaling behavior. The ZBP formed through spin exchange is a \textit{normal} Kondo peak with an energy scale characterized by $T_{\rm NK}$, whereas the peak formed without spin exchange is \textit{anomalous}, characterized by $T_{\rm AK}$.

This distinction is further supported by comparing the two temperatures, $T_{\rm NK}$ and $T_{\rm AK}$, with half the full width at half maximum (FWHM) of their respective ZBPs. According to reference~\cite{Cronenwett}, which investigated QPCs, half the FWHM of the zero-bias anomaly agrees well with the scaling temperature $T_{\rm NK}$ associated with $G_{\rm II}(T)$ in the conductance region above $0.7(2e^2/h)$. The reason why the region below $0.7(2e^2/h)$ is excluded has been discussed in reference~\cite{iop-qpc}.  

\begin{figure}[t] 
\centering
\includegraphics[width=3.0 in]{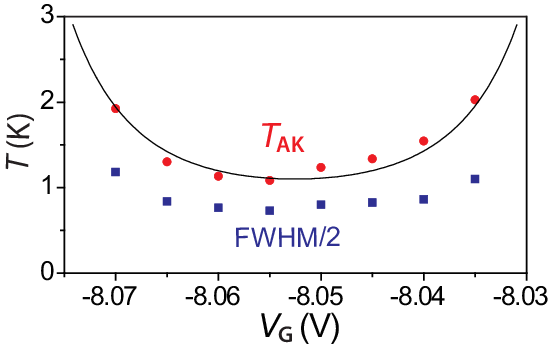}
\caption{$T_{\rm AK}$ data are obtained through scaling analysis of the temperature-dependent linear conductance in the leftmost sector of figure~\ref{fig5}(c), while $(1/2)$FWHM data are extracted from gate-voltage-dependent differential conductance line shapes measured in the same sector.}
\label{fig6}
\end{figure}

In contrast, the results of reference~\cite{Wiel} for QDSETs and those for CNT-based QDSETs shown in figure~\ref{fig6}, where $T_{\rm AK}$ are provided by the researchers who produced figure~\ref{fig5}, indicate that the scaling temperature $T_{\rm AK}$ associated with $G_{\rm I}(T)$ does \textit{not} coincide with half the FWHM of the ZBP. 

A notable study on the geometric crossover between QPC and QD~\cite{Heyder} provides additional insight: the distance between localized and reservoir spins in QD devices is considerably larger than in QPCs. A greater separation between the two spins forming a Kondo singlet weakens the Kondo coupling strength, effectively suppressing spin exchange. 
It is therefore understandable that the ZBP in the QDSET odd sector can form without spin exchange ($\gamma^L = \gamma^R = 0$).

\section{Conclusion}\label{sec:conclusion}

We performed a scaling analysis of the coherent peaks observed in QPCs and QDSETs, and found that these peaks can be classified into two groups according to their scaling functions.
(1) ZBPs observed in the odd sector of QDSETs~\cite{Goldhaber-PRL,Petit,Grobis,Wiel,Heersche} and all finite-bias coherent peaks adopt the scaling function 
$G_{\rm I}(T)$.
(2) ZBPs observed in the triplet state of the QDSET even sector~\cite{Roch} and in QPC devices~\cite{Cronenwett,Chen} adopt the scaling function 
$G_{\rm II}(T)$.

Furthermore, we revealed that the ZBPs in the QDSET odd sector and those in QPCs differ in both spin dynamics and scaling behavior. Specifically, the former peak arises from the combination of two coherent side peaks—which explicitly appear in QPC devices~\cite{Cronenwett,Sarkozy,Chen,Ren,Kristensen,DiCarlo}—and its half the FWHM does not coincide with the scaling temperature $T_{\rm AK}$ of $G_{\rm I}(T)$ (see reference~\cite{Wiel} and Fig.~\ref{fig6}).

From these findings, we conclude that the scaling temperature $T_{\rm AK}$ associated with the scaling function $G_{\rm I}(T)$ represents an \textit{anomalous} Kondo temperature, since in this case a Kondo singlet is formed between the localized spin and a reservoir spin, but spin exchange is deactivated.


\centerline{\bf ACKNOWLEDGMENTS}

 The author appreciates B.-K. Kim and M.-H. Bae of the Korea Research Institute of Standards and Science for providing unpublished data on differential conductance.

\appendix   
\section{Matrix elements}
\setcounter{section}{1} 

The matrix elements of the first term in equation~(\ref{reduced}) are given by
\begin{eqnarray}
\gamma^{L(R)}=\frac{\langle\sum_ki(\tilde{V}^Lc_{k\uparrow}^L+\tilde{V}^Rc^R_{k\uparrow})d^\dagger_{\uparrow}
		[j^{-L(R)}_{d\downarrow},j^{+L(R)}_{d\downarrow}]\rangle^{\rm sn}}
{\langle(\delta j^{-L(R)}_{d\downarrow})^2\rangle^{1/2}\langle(\delta j^{+L(R)}_{d\downarrow})^2\rangle^{1/2}}, \nonumber \\
\label{gamma-L}
\end{eqnarray}
\begin{equation}
\gamma^{LR}_S=\frac{\langle\sum_{k,k'}i(\tilde{V}^Lc_{k\uparrow}^L+\tilde{V}^Rc^R_{k'\uparrow})d^\dagger_{\uparrow}
		[j^{-L}_{d\downarrow},j^{+R}_{d\downarrow}]\rangle^{\rm sn}}
{\langle(\delta j^{-L}_{d\downarrow})^2\rangle^{1/2}\langle(\delta j^{+R}_{d\downarrow})^2\rangle^{1/2}}, 
\label{gamma-S}
\end{equation}
\begin{equation}
\gamma^{LR}_A=\frac{\langle\sum_{k,k'}i(\tilde{V}^Lc_{k\uparrow}^L+\tilde{V}^Rc^R_{k'\uparrow})d^\dagger_{\uparrow}
		[j^{-L}_{d\downarrow},j^{-R}_{d\downarrow}]\rangle^{\rm sn}}
{\langle(\delta j^{-L}_{d\downarrow})^2\rangle^{1/2}\langle(\delta j^{-R}_{d\downarrow})^2\rangle^{1/2}}, 
\label{gamma-A}
\end{equation}  
and
\begin{eqnarray}
U_{j^\mp}^{\nu}=i\frac{U}{2}\frac{\langle [n_{d\downarrow},j^{\mp\nu}_{d\downarrow}]
(1-2n_{d\uparrow})+j^{\mp\nu}_{d\downarrow}(1-2\langle n_{d\downarrow}\rangle)\rangle^{\rm eq}}
{\langle(\delta j^{\mp\nu}_{d\downarrow})^2\rangle^{1/2}}, \nonumber \\
\label{U} 
\end{eqnarray}
where $\tilde{V}^\nu$ is $V_{\rm G}$-dependent hybridization strength, 
$j^{-\nu}_{d\downarrow} = i{\tilde V}^\nu\sum_k(c^{\nu\dagger}_{k\downarrow} d_{\downarrow} - d^\dagger_{\downarrow} c^\nu_{k\downarrow})$, and 
$j^{+\nu}_{d\downarrow} = {\tilde V}^\nu\sum_k(c^{\nu\dagger}_{k\downarrow} d_{\downarrow} + d^\dagger_{\downarrow} c^\nu_{k\downarrow})$, 
with $c^\nu_{k\downarrow}$ denoting the $\nu$ reservoir and $\nu\in\{L,R\}$.
The superscript “eq” in \(U_{j^\mp}^{\nu}\) indicates that the expectation value is taken at equilibrium.
The detailed calculation procedure is provided in the appendix of reference~\cite{iop-qpc}. 

\section{Entangled-state tunneling}
\setcounter{section}{2} 
In \(\gamma^L\) of equation~(\ref{gamma-L}), for example, the commutator is expressed as
\begin{equation}
[j^{-L}_{d\downarrow},j^{+L}_{d\downarrow}]=2i\sum_{k\in
L}({\tilde{V}}^L)^2(c^{L\dagger}_{k\downarrow}d_\downarrow d^\dagger_\downarrow c^L_{k\downarrow}
-d^\dagger_\downarrow c^L_{k\downarrow}c^{L\dagger}_{k\downarrow}d_\downarrow).
\label{LL}
\end{equation}

Unidirectional entangled-state tunneling from the left to the right reservoir is formed by selecting four $\downarrow$-spin operators,
$c^{L\dagger}_{k\downarrow}d_\downarrow d^\dagger_\downarrow c^{L}_{k\downarrow}$, from equation~(\ref{LL}), 
and four $\uparrow$-spin operators: $c_{k\uparrow}^L d^\dagger_\uparrow$ from $(\tilde{V}^L c_{k\uparrow}^L+\tilde{V}^R c^R_{k'\uparrow})d^\dagger_{\uparrow}$ in equation~(\ref{gamma-L}), and 
$c_{k'\uparrow}^{R\dagger} d_\uparrow$ from the definition of the steady-state nonequilibrium expectation~\cite{iop-qpc}, 
\begin{equation} 
\langle 6 \, {\rm operators} \rangle^{\rm sn} := \langle\Psi_0|c_{k'\uparrow}^{R\dagger} (6 \, {\rm operators})  d_\uparrow |\Psi_0 \rangle,
\label{non-eq}
\end{equation}
where the operators \(d_\uparrow\) and \(c_{k'\uparrow}^{R\dagger}\) represent the initial and final events of a cyclic steady-state process, respectively, and 
the ground state \(|\Psi_0 \rangle\) describes a combined electron configuration of the left and right reservoirs at the Fermi level. 

The eight selected operators can be rearranged as 
$(d_\uparrow^\dagger c_{k\downarrow}^{L\dagger}) (c_{k\uparrow}^L d_\downarrow) (c_{k'\uparrow}^{R\dagger}d_\downarrow^\dagger) (c_{k\downarrow}^L d_\uparrow)$, 
which corresponds to the following sequence:
(left singlet formation) $\rightarrow$ (singlet co-tunneling) $\rightarrow$
(singlet formation on the right) $\rightarrow$ (spin exchange), as shown in the top row of figure~\ref{fig4}.

Meanwhile, the commutators in $\gamma^{LR}_{S}$ and $\gamma^{LR}_{A}$ from equations~(\ref{gamma-S}) and~(\ref{gamma-A}), respectively, are written as
\begin{eqnarray}
&&[j^{-L}_{d\downarrow},j^{+R}_{d\downarrow}]=i\sum_{k\in
L}\sum_{k'\in R}\tilde{V}^L\tilde{V}^R(c^{R\dagger}_{k'\downarrow}d_\downarrow d^\dagger_\downarrow c^L_{k\downarrow} \nonumber \\
&-&d^\dagger_\downarrow c^L_{k\downarrow}c^{R\dagger}_{k'\downarrow}d_\downarrow
+c^{L\dagger}_{k\downarrow}d_\downarrow d^\dagger_\downarrow c^R_{k'\downarrow}-d^\dagger_\downarrow c^R_{k'\downarrow}c^{L\dagger}_{k\downarrow}d_\downarrow),
\label{gamma-SS}
\end{eqnarray}
and 
\begin{eqnarray}
&&[j^{-L}_{d\downarrow},j^{-R}_{d\downarrow}]=-\sum_{k\in L}\sum_{k'\in R}\tilde{V}^L\tilde{V}^R(c^{R\dagger}_{k'\downarrow}d_\downarrow d^\dagger_\downarrow c^L_{k\downarrow} \nonumber \\
&-&d^\dagger_\downarrow c^L_{k\downarrow}c^{R\dagger}_{k'\downarrow}d_\downarrow 
-c^{L\dagger}_{k\downarrow}d_\downarrow d^\dagger_\downarrow c^R_{k'\downarrow}+d^\dagger_\downarrow c^R_{k'\downarrow}c^{L\dagger}_{k\downarrow}d_\downarrow).
\label{gamma-Anti}
\end{eqnarray}

The first terms in both expressions are identical and are used to construct unidirectional entangled-state tunneling from the left to the right reservoir as four 
$\downarrow$-spin operators. 
The four $\uparrow$-spin operators are the same as in the previous case. 
Rearranging the eight selected operators yields 
$(d_\uparrow^\dagger c_{k'\downarrow}^{R\dagger}) (c_{k\uparrow}^L d_\downarrow) (c_{k'\uparrow}^{R\dagger}d_\downarrow^\dagger)
(c_{k\downarrow}^L d_\uparrow)$, 
which represents the sequence:
(left singlet formation) $\rightarrow$ (singlet co-tunneling) $\rightarrow$
(new singlet formation on the right) $\rightarrow$ (singlet co-tunneling), 
as illustrated in the bottom row of figure~\ref{fig4}.

\end{document}